\documentclass[showpacs,preprint,amssymb]{revtex4}
\usepackage{graphicx}% Include Figure files
\usepackage{dcolumn}% Align table columns on decimal point
\usepackage{bm}% bold math
\def\be{\begin{equation}} 
\def\ee{\end{equation}}
\def\bea{\begin{eqnarray}} 
\def\eea{\end{eqnarray}}
\def\line{\hbox to \hsize}    
\def\frac #1#2{{#1\over #2}}

\def\sgn{{\rm sgn\,}}

\def \brak #1#2{{\langle#1\vert#2\rangle}}
 
\def\vev #1{{\langle #1\rangle}}
\def\1{\mbox{\bf 1}}

\begin{document}
%\draft %(only for revtex) 

\title{The Quantum Hydrodynamics of the Sutherland Model}

\author{ MICHAEL STONE}

\affiliation{University of Illinois, Department of Physics\\ 1110 W. Green St.\\
Urbana, IL 61801 USA\\E-mail: m-stone5@uiuc.edu}   

\author{DMITRY GUTMAN}

\affiliation{University of Florida\\ Department of Physics\\ PO Box 118440\\
 Gainesville, Florida 3261\\ 
E-Mail: gutman@phys.ufl.edu }

\begin{abstract}

We show that the  form of  the chiral condition found by  Abanov {\it et al\/}.\  in the quantum hydrodyamics of the Sutherland model arises because there are two distinct  inner products with respect to which   the chiral Hamitonian   hermitian,  but  only one with respect to which the full, non-chiral, Hamiltonian is hermitian.

\end{abstract}

\pacs{ 71.27.+a, 02.30.Ik, 05.30.-d, 71.10.Pm}

\maketitle

\section{Introduction}

It has long been understood that there is a close relationship  between the one-dimensional Luttinger-Thirring model \cite{luttinger} and the low energy edge modes  of  two-dimensional  Laughlin-state fractional quantum Hall  \cite{wen} fluids.  The two systems have very similar ground-state wavefunctions and the edge-particle correlation functions can be computed from the  Luttinger wavefunction \cite{stone-fisher}. The connection  exists because    
the non-commuting $x$ and $y$ coordinates of the lowest Landau level can be regarded  as the position and momentum co-ordinates of  a fluid of one-dimensional   fermions \cite{azuma}, and the boundary of the Hall fluid as its  Fermi surface.

If  we go beyond a linear approximation to the edge-state energies, the dimensional reduction  becomes both more interesting and more complicated. The Hall fluid may be described by  a  Chern-Simons matrix model \cite{susskind}. By adding boundary terms to the matrix model,   Polychronakos showed that  when a circular droplet of  quantum Hall two-dimensional electron fluid   is  held in place by a harmonic   $x^2+y^2$ potential, the $x$-axis projected  system  becomes \cite{polychronakos_chern} a  quantum Calogero  model \cite{calogero,calogero-sutherland} with the $y^2$ part of the   two-dimensional potential providing the non-relativistic kinetic energy and the $x^2$ part providing a one-dimensional harmonic   confining potential.  When   the Hall fluid is  confined   by a $y^2$ potential to a finite strip  with $x$-periodic boundary conditions, the one-dimensional system becomes \cite{polychronakos_cylinder} the periodic Sutherland model \cite{sutherland1,sutherland2}. The ground-state wavefunction and low lying excitations still coincide with those of the Luttinger model, but  the   higher exited states are more complicated.  There is a one-to-one mapping of the eigenstates of the harmonically  confined quantum Hall systems onto the eigenstates of the  Calogero-Sutherland models \cite{hellerman}, and this mapping descends to the soliton and small amplitude wave solutions  of the continuum classical hydrodymamics of the Calogero-Sutherland fluid \cite{polychronakos_soliton}. 

One curious feature of this mapping is that we are  looking at the  two-dimensional quantum Hall fluid sideways-on, and  see  both its  near and far  edges  superimposed. Although the two  boundaries 
have their  own independent edge modes  that  move in opposite directions, it is not easy  to make a clean left-right separation in the projected one-dimensional  quantum hydrodynamics \cite{polychronakos_soliton}.  Recently, however, Abanov {\it et al.\/}.\  \cite{abanov} showed that  the complicated non-local hydrodynamic equations were much simplified when expressed in terms of   a dynamical field  $u(z,t)$ that  lives on  a Schottky double constructed by gluing two copies of a non-compact  complex plane together along their boundaries.  This  ingenious  reformulation, which depends on an unusual form of Hilbert transform,   enabled them to find  a condition linking the density and velocity under which only unidirectional motion is excited \cite{bettelheim}.
The  definitions   of the field and unconventional Hilbert transform in \cite{abanov} are   not at all obvious, however, and the way in which the chiral condition works  seems almost magical. 

The present paper is devoted to an alternative  formulation of the quantum hydrodynamics  that  avoids  the Schottky-double contour integrals that are the key element  in  \cite{abanov,bettelheim}.  The non-obvious form of the chiral condition arises because there are    two distinct inner products with respect  to which the Jack-polynomial eigenfunctions  of the Sutherland model are mutually orthogonal \cite{macdonald,awata}. The first of   these is  the one most often met with in the literature of symmetric functions. The second  is the one that arises  from the quantum mechanics.  The chiral version of the Sutherland model is hermitian with respect to both these inner products. The full, non-chiral, version is hermitian only with respect to the second inner product. The mysterious  terms that appear in   \cite{abanov,bettelheim} are precisely the corrections  required to make the ``natural''  chiral  fields into  operators  that are hermitian with respect to the  second product. 

In section \ref{SEC:sutherland} we provide a brief account  of the Sutherland model. In section \ref{SEC:collective}  we review  the application of the collective field formalism \cite{sakita_book} to this model,  and introduce the two inner products. In section  \ref{SEC:inner}  we show how the difference between the two products manifests itself in  the quantum hydrodynamics, and in section \ref{SEC:ambichiral} how this difference is the origin of the complications in the chiral decomposition.

\section{The Sutherland  Model}
\label{SEC:sutherland}

We begin with a short review of the Sutherland model \cite{sutherland1}.
This consists of  $N$ particles moving on the unit circle with   Hamiltonian
\be
H_{\rm Sutherland} = -\frac 12 \sum_{i=1}^N \frac{\partial^2}{\partial \theta_i^2} +\frac{1}{4}\sum_{i<j}\frac{\lambda(\lambda-1)}{\sin^2(\theta_i-\theta_j)/2}.
\ee
We will restrict ourselves to the parameter range $\lambda >1$, where  the inter-particle interaction is repulsive. 
The potential  is sufficiently  singular  that  tunneling does not occur, and  the particles retain their original order around the circle. The  exchange  statistics of the particles  are therefore   unimportant,  but we will usually think of them as being fermions, as this  is their natural description in the limit $\lambda \to 1_+$.

The  ground state wavefunction and energy  may be found   by means of a ``supersymmetric quantum mechanics'' trick.
We set 
\be
\Delta =\prod_{i<j} 2\sin (\theta_i-\theta_j)/2,
\ee
and make use of     
the addition formula
\be
\cot(x-y)\cot(y-z)+\cot(y-z)\cot(z-x)+\cot(z-x)\cot(x-y)=1
\ee
to  write 
\bea
H&\equiv& H_{\rm Sutherland} - \frac {\lambda^2}{24}N(N^2-1)\nonumber\\
&=& -\frac 12 \sum_i \frac{\partial^2}{\partial \theta_i^2} +\frac{1}{4}\sum_{i<j}\frac{\lambda(\lambda-1)}{\sin^2(\theta_i-\theta_j)/2} - \frac {\lambda^2}{24}N(N^2-1)\nonumber\\
&=& \frac 12 \sum_i\left(-\frac{\partial}{\partial \theta_i} -\frac \lambda 2\sum_{j\ne i} \cot(\theta_i-\theta_j)/2\right) \left(\frac{\partial}{\partial \theta_i} -\frac \lambda 2\sum_{j\ne i} \cot(\theta_i-\theta_j)/2\right) \nonumber\\
&=& \frac 12 \sum_i \left(-\frac{1}{\Delta^\lambda}\frac{\partial}{\partial \theta_i} \Delta^\lambda\right)
\left({\Delta^\lambda}\frac{\partial}{\partial \theta_i} \frac{1}{\Delta^\lambda}\right)\nonumber\\
&=&\frac 12 \sum_{i=1}^N Q_i^\dagger Q_i.
\eea
It is now clear that    
\be
\Psi_0=\Delta^\lambda(\theta) = \left(\prod_{i<j} 2\sin (\theta_i-\theta_j)/2.\right)^\lambda 
\ee
satisfies $Q_i\Psi_0=0$ for all $i$,  and  is the unique  zero-energy   eigenfunction of $H$. 
The wavefunction $\Psi_0$ is therefore the ground state  of the  Sutherland model,  and the ground-state energy is 
\be
E_0= \frac {\lambda^2}{24}N(N^2-1).
\label{EQ:GS_energy}
\ee
This energy reduces for $\lambda=1$  to the energy of an $N$-particle  Fermi sea  obeying       periodic  boundary conditions when $N$ is odd and antiperiodic  boundary conditions when $N$ is even.

Now we seek wavefunctions of form $\Psi=\Delta^\lambda \Phi(z_1,\ldots,z_N)$, {\it i.e.\/}\  functions such that $H'\Phi=E'\Phi$, where
\bea
H' 
&=&-\frac 12 \sum_{i=1}^N \frac{1}{\Delta^{2\lambda}} \frac{\partial}{\partial \theta_i} \Delta^{2\lambda}\frac {\partial}{\partial \theta_i}\nonumber\\
&=&-\frac 12 \sum_i \frac{\partial^2}{\partial \theta_i^2} -\frac \lambda 2\sum_{i,j\ne i}  \cot(\theta_i-\theta_j)/2\frac{\partial}{\partial \theta_i}\nonumber\\
&=&-\frac 12 \sum_i \frac{\partial^2}{\partial \theta_i^2} -\frac \lambda 2 \sum_{i<j}  \cot(\theta_i-\theta_j)/2 \left( \frac{\partial}{\partial \theta_i}- \frac{\partial}{\partial \theta_j}\right)\nonumber\\
&=& \frac 12 \sum_i D_i^2 +\frac \lambda 2 \sum_{i<j}  \frac{z_i+z_j}{z_i-z_j} (D_i-D_j).
\eea
In the last  line we have set  $z_i=\exp\{i\theta_i\}$ and $D_i = z_i\partial/\partial z_i= -i\partial/\partial \theta_i$.  
The Hamiltonian $H'$ will be hermitian with respect to the ``$\lambda$-Sutherland''  inner product
\be
\brak{\Phi_1}{\Phi_2}_{\rm Sutherland}=\frac 1{N!} \int_0^{2\pi}\ldots \int_0^{2\pi}\prod_{i=1}^N \frac{d\theta_i}{2\pi}  |\Delta|^{2\lambda} \Phi_1^*\Phi_2. 
\ee

We initially  only consider $\Phi(z_1,\ldots,z_N)$ that are symmetric polynomials in the $z_i$. These  describe excitations near the $k\sim k_f$ Fermi point. Later we will   worry about the $z_i^{-1}$'s that can be used for the $k\sim {-k_f}$  Fermi point.

Sutherland \cite{sutherland2} considered, in particular, the  action of $H'$ 
on the 
monomial symmetric functions
\be
m_{\{\alpha\}}(z) = \sum z_{i_1}^{\alpha_1} z_{i_2}^{\alpha_2}\ldots z_{i_N}^{\alpha_N},
 \ee
where 
\be
\{\alpha\}\equiv  \{\alpha_1, \alpha_2,  \ldots, \alpha_N\}, \quad  \alpha_1\ge \alpha_2\ge \cdots \ge \alpha_N,
\ee
and 
the sum is over all permutations of the labels $i$ that result in distinct monomials.

 We can represent the integer sequence $\alpha_i$  by a Young  (or Ferrars) diagram with $\alpha_1$ boxes in the first row, $\alpha_2$ in the second, and so on, and think of it as a partition of the integer $|\{\alpha\}|\equiv \alpha_1+\alpha_2+\ldots$.
We usually order partitions   in reverse lexographic order, in which 
$\{\alpha\}> \{\beta\}$  if the first non zero difference $\alpha_i-\beta_i$ is positive. This is a {\it total\/}  ordering: given two partitions one is greater than the other, or they are equal.
An alternative  ordering is {\it dominance ordering\/} in  which $\{\alpha\}\succeq \{\beta\}$ if 
\be
\sum_{i=1}^n \alpha_i \ge \sum_{i=1}^n \beta_i,\quad \forall  n>0.
\ee
Dominance is only a {\it partial\/}  order (as is  set inclusion) in that not all partitions are comparable. If 
$\{\alpha\}\succ  \{\beta\}$, however, then $\{\alpha\}> \{\beta\}$. Sutherland showed that when  the $m_{\{\alpha\}}$ are taken as a basis, then  $H'$ is represented by a matrix that is  upper triangular with respect to dominance order. 
The eigenvalues of $H' $are  therefore the diagonal elements of this upper-triangular matrix. His  result \cite{sutherland2} is that the eigenvalues of $H_{\rm Sutherland}$ can be written as  
\be
E_{\{\alpha\}} = E_0+E'_{\{\alpha\}}=\frac 12 \sum_i \xi_i^2.
\ee
where the {\it pseudomomenta\/}  $\xi_i$ are 
\be
\xi_i = \alpha_i +\lambda k_i^0,
\ee
with $k_i^0$ being the momenta of the Fermi sea of free fermions. The ground state has all the $\alpha_i=0$, and so we recover the formula (\ref{EQ:GS_energy}) for the ground-state energy.

The polynomial eigenfunctions  of $H'$ are the {\it  Jack symmetric functions}  $J_{\{\alpha\}}(z)$. They  can, in principal be found by using $\brak{\dots}{\ldots}_{\rm Sutherland}$ to 
apply the  Gramm-Schmidt procedure to the reverse-lexographically-ordered monomial symmetric functions. It is a non-trivial result \cite {macdonald} that the only subtractions appearing in the orthogonalization process  involve dominance-ordered $m_{\{\beta\}}$.
 Thus we obtain  
 \be
J_{\{\alpha\}}(z)= m_{\{\alpha\}}(z) +\sum_{\{\beta\} \prec \{\alpha\}} K_{\{\alpha\}\{\beta\}} m_{\{\beta\}}(z),
\ee 
and this condition, together with  $\brak{J_{\{\alpha\}}}{J_{\{\beta\}}}_{\rm Sutherland}=0$ when $\{\alpha\}\ne \{\beta\}$, serves to define  the Jack functions  uniquely.

When $\lambda=1$, the Jack polynomials reduce to the  Schur symmetric functions and the  coefficients $K_{\{\alpha\}\{\beta\}}$ become  Kostka numbers. Both the Schur and Jack functions are zero whenever the length $l(\{\alpha\})$ of the partition (the number of non-zero rows in the Young diagram) exceeds $N$.

\section{Collective fields}  
\label{SEC:collective}

We wish to describe the low energy and low momentum excitations of the Sutherland chain in terms of fluctuations in the particle density. 
We therefore  change variables from $z_1,\ldots z_N$ to $p_1,\ldots,p_N$, where $p_n=\sum_i z_i^n$ are the Newton power-sum symmetric functions, and simultaneously  the positive-momentum Fourier components of the density $\rho(\theta)=\sum_i^N \delta (\theta-\theta_i)$:
\be
p_n= \int_0^{2\pi}   e^{in\theta}\rho(\theta)\,d\theta.
\ee
We find that 
\be
D_i=z_i\frac{\partial}{\partial z_i} =z_i \sum_{n=1}^N \frac{\partial p_n}{\partial z_i}\frac{\partial}{\partial p_n} = \sum_{n=1}^N n z_i^n \frac{\partial}{\partial p_n},
\ee
and from this obtain  
\be
\sum_i D_i^2 = \sum_{n=1}^N n^2 p_n \frac{\partial}{\partial p_n} + \sum_{m,n=1}^N nm\,p_{n+m}\frac{\partial}{\partial  p_n}\frac{\partial}{\partial  p_m}.
\label{EQ:d2}
\ee
We now use 
\be
(z_i+z_j)\frac{z_i^n-z_j^n}{z_i-z_j} = z_i^{n} +2z_i^{n-1}z_j+\cdots +2z_iz_j^{n-1}+z_j^{n}
\ee
and, after some careful tracking of duplicated and omitted terms,   obtain
\be
\frac{\lambda}{2} \sum_{i \ne j}  \frac{z_i+z_j}{z_i-z_j} (D_i-D_j)
=\lambda \sum_{m,n= 1}^{n+m\le N} (m+n) p_{m}p_{n}\frac{\partial}{\partial p_{m+n}} +  \lambda \sum_{n=1}^N n(N-n)p_n\frac{\partial}{\partial p_n}.
\label{EQ:restricted}
\ee
Note that neither $m$ nor $n$ is allowed to be  zero in the first sum on the right. The excluded  terms---those at the ends,   without  2's,  in the $z$ series---appear as   the $nN$ part of the second sum. The $- n^2$ part  arises  from the restriction that $i$ cannot equal $j$.  

The 
$n+m\le N$ constraint in the first sum   on the right in (\ref{EQ:restricted}) is natural because only $p_n$'s with $n\le N$ are algebraically independent,  so the wavefunction, when expressed in terms of the $p_n$'s,  should not contain $p_n$'s with $n>N$. Correspondingly, any   $p_{n+m}$ with $n+m>N$ generated by an application of the operator in  (\ref{EQ:d2})  should, in principle,  be re-expressed in terms of $p_n$'s with $n\le N$ by means  of the Newton-Girard relations.
It is, however, not unreasonable to ignore these issues in the collective field formalism. This is because we are ultimately interested in taking a thermodynamic limit in which we simultaneously rescale the mass of particles and  the circumference of the circle so as to let $N\to \infty$ while keeping the physical density and non-relativistic dispersion fixed.

If  we ignore the $n+m\le N$ constraint and allow the sums to extend to infinity,  we have 
\be
2H'= \sum_{n,m=1}^\infty \left(nmp_{n+m} \frac{\partial}{\partial p_n} \frac{\partial}{\partial p_m} +\lambda(m+n)p_mp_n  \frac{\partial}{\partial p_{m+n}}\right) +\sum_{n=1}^\infty\left((1-\lambda)n^2+\lambda nN\right)p_n\frac{\partial}{\partial p_{n}}.
\label{EQ:Hprime}
\ee
Now $H'$ should  be Hermitian, and the right-hand-side of  (\ref{EQ:Hprime})  is manifestly so if    
\be
p_n^\dagger = \frac n\lambda  \frac{\partial}{\partial p_n}. 
\ee
We know, from standard chiral bosonization \cite{stone_bosonization}, that this identification is correct for $\lambda=1$.
Accepting the identification for general $\lambda$, we can evaluate the  inner product
\bea
\brak{p_{\{\alpha\}}}{p_{\{\beta\}}}_{\rm Jack}&\equiv& \brak{p_1^{m_1}\cdots p_N^{m_N}} {p_1^{n_1}\cdots p_N^{n_N}}_{\rm Jack} \nonumber\\
&=&\brak{1}{(p_N^\dagger)^{m_N}  \cdots (p_1^\dagger)^{m_1} p_1^{n_1}\cdots p_N^{n_N}}_{\rm Jack} \nonumber\\
&=&\delta_{\{\alpha\}\{\beta\}} \lambda^{-l\{\alpha\}} \prod_{i=1}^N m_i!\, i^{m_i}.
\label{EQ:jack_product}
\eea
Here we are using a slightly different parametrization of the  partitions: $\{\alpha\}\equiv \{1^{m_1}2^{m_2}\ldots\}$, where the integer  $m_i$ is the number of rows in the Young diagram of $\{\alpha\}$ containing  $i$ boxes, and  so  the length  of the partition is given by $l(\{\alpha\})\equiv m_1+m_2+\cdots$.  The  expression  (\ref{EQ:jack_product}) for the inner product of the $p_{\{\alpha\}}$  defines what we will call the  ``$\lambda$-Jack'' inner product. 
It  can be expressed in  Bargmann-Fock integral  form as
\be
\brak{F(p)}{G(p)}_{\rm Jack}= \int \prod_{n=1}^{\infty} \left(\lambda \frac{ d^2 p_n}{\pi n}\right) [F(p)]^*G(p) \exp\left\{-  \lambda\sum_{n=1}^{\infty} \frac 1 n p_n^*p_n\right\}, 
\ee where
$d^2 p_n = d[{\rm Re\,}p_n] \,d[ {\rm Im\,}p_n]$ and each  integration is over the entire complex $p_n$ plane. 

Macdonald \cite{macdonald} uses this new inner product  to define  the Jack polynomials by  again applying the Gramm-Schmidt procedure to the monomial  symmetric functions $m_{\{\alpha\}}(z)$. 
Now the  $\lambda$-Sutherland and the $\lambda$-Jack inner products  are in general different. They    only coincide when $\lambda=1$ (this is the miracle behind conventional bosonization) or when $N$ is  infinite.
Remarkably, however, the  Gramm-Schmidt procedure yields the same polynomials whichever product is used. This is because the Jack polynomials  are  mutually orthogonal with respect to {\it both\/}  inner products, although   their  norms differ.

If, for $n>0$,  we set $j_n = p_n $, $j_{-n}= j^\dagger_{n}= p^\dagger_n$  and $\nu=\lambda^{-1}$,  we have the filling-fraction $\nu$   chiral  algebra
\be
[j_n, j_m]=\nu  m\,\delta_{n+m,0}.
\ee
We  also  set $j_0=N/2$, anticipating that the other $N/2$ will go in the left-going current. 
In position space
 \be
 j(\theta)= \frac 1{2\pi} \sum_{n=-N}^N j_ne^{-in\theta},
 %\quad j_0=N/2, 
\ee
and   the current algebra  becomes  
 \be
[ j(\theta),j(\theta')] =-\frac{i\nu}{2\pi}\delta'(\theta-\theta'),
 \ee
 which is the familiar  right-going current commutator, at least   at $\nu=1$. 

In terms of the current components $j_n$ we can write 
\be
2H' =\lambda^2  \sum_{n,m=1}^\infty\left(j_{n+m} j_{-n}j_{-m}+j_n j_m j_{-n-m}\right)  +\lambda \sum_{n=1}^\infty\left((1-\lambda)n+\lambda N\right)j_n  j_{-n},
\ee
 which is manifestly Hermitian with respect to the $\lambda$-Jack inner product, and normal-ordered.
 In position space the cubic terms in $2H'$ become  
 \be
4\pi^2\int_0^{2\pi} \frac{\lambda^2}{3} j(\theta)^3 d\theta, 
 \ee
 where  normal-ordering is to be understood.
 The  quadratic terms   can be written as an integral of a  periodic Hilbert transform
 \be
[\varphi(\theta)]_{\rm H}\stackrel{\rm def}{=} \frac 1{2\pi} P\int_0^{2\pi} \varphi(\theta')\cot\left(\frac{\theta-\theta'}{2}\right)\,d\theta' 
\nonumber\\
\ee
for which   $(e^{-in\theta} )_{\rm H}= i\,\sgn(n) e^{-in\theta}$.
 We find that  
 \bea
2 H' &=& 4\pi^2 \int_0^{2\pi}\left\{ \frac{\lambda^2}{3}j(\theta)^3 -i a j(\theta)\partial_\theta(j_+(\theta)-j_-(\theta))\right\} d\theta,\nonumber\\
 &=&4\pi^2 \int_0^{2\pi}\left\{ \frac{\lambda^2}{3}j^3 - a j\partial_\theta j_{\rm H}\right\} d\theta. 
  \label{EQ:chiralform}
  \eea
 Here 
 \be
 a= \lambda(\lambda-1)/4\pi,
\ee
 and $j_+ $ is the part of $j$ with $j_n$, $n>0$, and similarly $j_-$ has $j_n$ with $n<0$. 
 
 %Note that with our  definition $(e^{-in\theta} )_{\rm H}= i\,\sgn(n) e^{-in\theta}$, and the inverse Hilbert transform is 
 %\be
%\varphi(\theta)= -\frac 1{2\pi} P\int_0^{2\pi} [\varphi(\theta')]_{\rm H}\cot\left(\frac{\theta-\theta'}{2}\right)\,d\theta' 
%+\frac 1{2\pi} \int_0^{2\pi} \varphi(\theta')\,d\theta'.
%\ee

 The resulting classical (where $\lambda(\lambda-1)\to \lambda^2$, because the ``1'' is really an $\hbar$) equation of motion is of   Benjamin-Ono form  
 \be
 \partial_\tau j+ j\partial_\theta j - \beta\,\partial^2_{\theta\theta}j_{\rm H}=0,
 \ee
 where $\tau = 2\pi \lambda t$, and $\beta=1/4\pi$.  Seen from a frame moving at the speed of sound $c=\pi \lambda \rho_0$--- so as to  remove the convective effect of the constant background $\vev{j}=\rho_0/2$ ---the   Benjamin-Ono equation on the infinite  line has  a right-going solition solution  
 \be
 j(x,t)-\vev{j}= \frac{4U}{\beta^{-2} U^2[x- U\tau]^2+1}.
 \ee
 Here $2\pi \lambda U= (v_{\rm soliton}-c) $ must be positive, so the solitons always travel faster than the speed of sound \footnote[1]{
 The positivity condition on $U$ stems from the $|a|$ in 
$
\left( \frac 1{x^2+a^2}\right)_{\rm H}=\frac 1{|a|} \frac {x}{x^2+a^2}
$.}.
 The excess charge carried by the solition is  
 \be
 \int_{-\infty}^{\infty} ( j(x,t)- \vev{j})\,dx= 4\pi \beta=1.
 \ee
 This solution is close to, but not identical with, the soliton solution for the continuum approximation to the classical Calogero model found by Polychronakos \cite{polychronakos_soliton}. The difference is that  Polychronakos' solitions can travel both to the left and right, and  the width of his soliton is
 $\lambda c /[v_{\rm soliton}^2-c^2]$. The present   soliton width is  $\lambda/[2(v_{\rm soliton}-c)]$.  The two widths coincide, however,  when  $v_{\rm soliton}-c$ is small compared to $c$, {\it i.e.\/}\ when the excitation momentum  is small compared to the  distance between the left and right Fermi surfaces.

\section{The inner products and  the collective-field measure} 
\label{SEC:inner}

The Jack polynomials form an orthogonal, but not orthonormal, basis for the symmetric functions with respect to  both the Jack and Sutherland inner products.

We have  \cite{macdonald}
 \be
\brak{J_{\{\alpha\}}}{J_{\{\alpha\}}}_{\rm Sutherland}= \prod_{1\le i<j\le N}\frac{\Gamma(\xi_i-\xi_j+\lambda)\Gamma(\xi_i-\xi_j-\lambda+1)}{\Gamma(\xi_i-\xi_j)\Gamma(\xi_i-\xi_j+1)},
\ee
where the $\xi_i$ associated with the partition $\{\alpha\}$ are  
\be
\xi_i =\alpha _i+\lambda k_i^0
\ee
are the pseudomomenta,in terms of which the Sutherland energy eigenvalue is 
$$
E_{\{\alpha\}}= \frac 12 \sum_{i=1}^N \xi_i^2.
$$
The Jack  product, on the other hand,  gives   \cite{macdonald}
\be
\brak{J_{\{\alpha\}}}{J_{\{\alpha\}}}_{\rm Jack}= \prod_{s\in\{\alpha\}} \frac{a(s)+\lambda l(s) +1}{a(s) +\lambda l(s) +\lambda}.
\ee
Here $s$ labels  a box in the Young  diagram of the partition $\{\alpha\}$, and $a(s)$ and $l(s)$ are respectively the {\it arm length\/} (the number of boxes to the right of $s$) and {\it leg length\/} (the number of boxes below $s$) of $s$.

The relation between the two norms is   \cite{macdonald}
\be
\brak{J_{\{\alpha\}}}{J_{\{\alpha\}}}_{\rm Sutherland}= C_N\brak{J_{\{\alpha\}}}{J_{\{\alpha\}}}_{\rm Jack} \prod_{s\in\{\alpha\}} \frac{\lambda N+a'(s)-\lambda l'(s) }{\lambda N+a'(s) +1 -\lambda (l'(s)+1)},
\label{EQ:colength}
\ee
where $a'(s)$ and $l'(s)$ are respectively the {\it arm co-length\/} (the number of boxes to the left of $s$) and {\it leg co-length\/} (the number of boxes above $s$) of $s$, 
and 
% \cite{dyson_conjecture,wilson_proof}
\be
C_N=\brak{1}{1}_{\rm Sutherland}=\frac 1{N!}  \int_0^{2\pi}\ldots \int_0^{2\pi}\prod_{i=1}^N \frac{d\theta_i}{2\pi}  |\Delta|^{2\lambda} =\frac 1{N!} \frac{\Gamma(1+\lambda N)}{[\Gamma(1+\lambda)]^N}.
 \ee
Inspection of (\ref{EQ:colength}) shows that  scaled product $C_N^{-1}\brak{\ldots}{\ldots}_{\rm Sutherland}$ will  coincide with the Jack product  when $\lambda=1$ or when $N\to \infty$ with all  $a'(s)$ and $l'(s)$ remaining finite. 

The source of the  difficulty in decoupling the left and right Fermi-surface physics is that scaled Sutherland product need  {\it not\/}  coincide with the Jack product when $N\to \infty$ and at the same time the number of rows or columns in the Young diagram remains  $O(N)$.  The former  is exactly the situation when we when we seek to  describe excitations near the left-hand Fermi surface.
We  do this exploiting the identity
\be
(m_{\{1^N\}})^pJ_{\{\alpha_1,\alpha_2,\ldots, \alpha_N\}}(z) = J_{\{\alpha_1+p,\alpha_2+p,\ldots, \alpha_N+p\}}(z),
\ee
where $m_{\{1^N\}}(z)=z_1z_2\cdots z_N$,
to add $p$ columns of $N$ boxes on the left of the Young diagram representing  the Sutherland eigenstate. This operation \cite{awata} corresponds to a Galilean boost  in which each of the $N$ particles   is given an additional $p$ quanta of momentum. If $p$ is made large enough, all the particle momenta can be made positive.  The Sutherland inner product (but not the Jack  product) is invariant under such boosts.  We can   create  negative-momentum excitations near  the left-hand Fermi surface by removing boxes near the bottom of the, now $N$-row  deep, Young diagram. This means that   the left-most pseudo-momenta were  not boosted quite as far as the others. When the boost is undone by removing the added columns,  we are left with a Young diagram with some negative-length rows. The corresponding  $H'$ eigenfunction is now a rational function rather than a polynomial, but it can be written as a   conventional Jack polynomial multiplied by a negative power of $m_{\{1^N\}}$.   The Jack and Sutherland products will not coincide for such ambichiral states.

To understand the consequences of this  difference between the Sutherland and Jack products in the collective field language, we begin by  exploring  how it is that these rather differently defined products  become equal in the large-$N$ chiral case.

If $|z_i|=1$, and $|\mu|<1$ is a  convergence factor  inserted to make  the  logarithmic series converge, we have 
 \bea
 \exp\left\{-\lambda\sum_{n=1}^\infty \frac 1 n \mu^n p^*_np_n\right\}
&=&
 \prod_{i, j}(1-\mu z_i^*z_j)^{\lambda}\nonumber\\
 &=& \prod_{i<j}(1-\mu z^*_i z_j)^\lambda(1-\mu)^{N\lambda}  (1-\mu z^*_jz_i)^\lambda  \nonumber\\
 &= & \prod_{i<j} (z_i-\mu z_j)^\lambda(z_i^*-\mu z^*_j)^\lambda (1-\mu)^{N\lambda}
\nonumber\\
&\to & 
 \left|\Delta(z)\right|^{2\lambda}(1-\mu)^{N\lambda}, \quad \hbox{as $\mu\to 1_-$}.  
 \eea
 We see that the  explicit weights in the Sutherland  and Jack   products  are in some sense  proportional, but the constant of proportionality diverges to zero as $\mu\to 1$. 
 
We do not need a convergence factor  in  
 \be
 \sum_{n=-\infty}^{\infty} \frac 1{|n|} e^{in\theta}= -2\ln| 2 \sin (\theta/2)|,
 \ee
 and so with $\rho(\theta)= \frac 1{2\pi} \sum p_n e^{-in\theta}$ we have
 \be
\exp\left\{-\lambda  \sum_{n=1}^{\infty} \frac 1 n p_np_{-n}\right\}=\exp \left\{ \lambda \int_0^{2\pi} \int_0^{2\pi} \rho(\theta)\rho(\theta') \ln\left|2 \sin\left(\frac{\theta-\theta'}{2}\right)\right|d\theta d\theta'\right\}.
\ee
An additive  constant $p_0$ in $\rho$ does not contribute to the right-hand side   because the kernel integrates to zero.
The singularity in the integrand is integrable. What does this mean for the divergent ``$i=j$'' factors in the exact product?  Should the integral contain a   counterterm to remove them?
The appropriate replacement is \cite{dyson1}
\be
\prod_{i,<j}|z_i- z_j|^{2\lambda}\simeq C\exp  \left\{ \lambda \int_0^{2\pi}  \rho(\theta) \ln \rho(\theta)\, d\theta + \lambda \int_0^{2\pi} \int_0^{2\pi} \rho(\theta)\rho(\theta') \ln\left|2 \sin\left(\frac{\theta-\theta'}{2}\right) \right|\,d\theta d\theta'\right\}.
\label{EQ:coulomb}
\ee
The first term subtracts a $\ln( \hbox{interparticle spacing})$  self-energy for each particle, and    is consistent with the observation   that when the $z_i$ are equally spaced round the unit circle we 
\be
\prod_{i<j}|z_i- z_j|^{2\lambda}\to  (N^{N/2})^{2\lambda}=\exp\left\{\lambda N\ln N\right \}. 
\ee
In  a  ``coulomb gas'' interpretation  
the first term in the exponent  in (\ref{EQ:coulomb}) computes the microscopic internal  energy  of the uniform  gas, and the second accounts for the electrostatic energy  due to macroscopic deviations from uniformity.

In addition to expressing the $|\Delta|^{2\lambda}$ weight in terms of the particle density, we  need to compute  the Jacobian of the transformation from the $z_i$ to the $p_n$. This change of variable  is conceptually subtle. The map $(z_1,\ldots ,z_N)  \to (p_1,\ldots, p_N)$  is not  invertible:  each of  the  $z_i$ has unit modulus, whilst  in the Bargmann-Fock integral  the  $p_n$ are general  complex numbers.   An arbitrary set  of  $p_n$  will not arise from  from   $z_i$ with $|z_i|=1$.  However, as the $z_i$ move on their unit circles, each $p_n$ moves as the  endpoint of an  $N$-step random walk in the complex 
 plane with $\vev{|p_n|^2}=N$. 
 By the central limit theorem, therefore,  each $p_n$  has large-$N$  probability density
 \be
 P(p_n) =  (N\pi)^{-1} e^{-|p_n|^2/N}.
 \ee
 It is natural to   conjecture that as $N\to \infty$ the map $z_i\to z_i^n$ so scrambles the directions of the individual $z_i^n$ steps that their  sums $p_n=\sum_i z_i^n$ become  {\it independent\/}  random variables with joint probability density 
 \be
P(p_1, p_2,\ldots ) \propto  \exp\left\{ -\frac 1 N\sum_{n=1}^{\infty} p_np_{-n}\right\}= \exp\left\{-\frac 1{2\rho_0} \int_0^{2\pi} (\rho'(\theta))^2\, {d\theta} \right\}.
\label{EQ:jpdf}
\ee
Here $\rho'=\rho-\rho_0$ and $\rho_0=N/2\pi$.
 As $N$ becomes large this distribution becomes uniform on the scale of the early ($n\ll N$) exponentials in  the $\lambda$-Jack  Bargmann-Fock interal 
 and so  the low-momentum  integration measures in the Sutherland  and Jack products  are also proportional--- despite one integration domain having twice the dimension as the other. (In other words the large-$N$ image of the real  $N$-torus is dense in ${C}^N.$) 
 
 The integration measure will not appear  uniform if   applied to  wavefunctions containing $p_n$'s with $n=O(N)$. In this case we need a more accurate  formula.
Jevicki shows   \cite{jevicki1} that our  conjectured probability density (\ref{EQ:jpdf})   is but the first term in a systematic  expansion in powers of $1/N$:
 \be
P(p_1, p_2,\ldots ) \propto \exp\left\{\int_0^{2\pi} \left[ - \frac 1{2\rho_0}{\rho'}^2\ +\frac 1{6\rho_0^2} {\rho'}^3 - \frac 1{12\rho_0^3} {\rho'}^4+\cdots \right]{d\theta} \right\}.
\ee
Now  we observe that 
\be
\int_0^{2\pi}\left[- \frac 1{2\rho_0}{\rho'}^2\ +\frac 1{6\rho_0^2} {\rho'}^3 - \frac 1{12\rho_0^3} {\rho'}^4+\cdots\right]\,d\theta =
\int_0^{2\pi}\left[ \rho_0\ln\rho_0-(\rho_0+\rho')\ln(\rho_0+\rho')\right]\,d\theta,
\ee
and so surmise   that  
\be
P(p_1, p_2,\ldots ) \propto \exp\left\{- \int_0^{2\pi} \rho\ln\rho\, d\theta\right\}.
\ee
To verify  this  conjecture,   we can proceed as follows: 
we want to find the measure  $P[p_n]$ such that
\be
 \int_0^{2\pi}\ldots \int_0^{2\pi}\prod_{i=1}^N \frac{d\theta_i}{2\pi}F\left(\sum_i e^{in\theta_i}\right) = \int \prod_{n=1}^\infty dp_n   P[p_n] F(p_n).
\ee
Let $\lambda(\theta)= \sum_n \lambda_n e^{in\theta}$, and, as usual, $\rho(\theta) =\frac 1{2\pi} \sum_n p_ne^{-in\theta}$.
Thus
\bea
P[p_n]&=& \int_0^{2\pi}\ldots \int_0^{2\pi}\prod_{i=1}^N \frac{d\theta_i}{2\pi} \prod_{n=1}^\infty\delta\left(-p_n+ \sum_i e^{in\theta_i} \right)\nonumber\\
&=&  \int_0^{2\pi}\ldots \int_0^{2\pi}\prod_{i=1}^N \frac{d\theta_i}{2\pi}\int \prod_{n=1}^\infty d\lambda_n \exp\left\{ \lambda_n\left(-p_n+ \sum_i e^{in\theta_i}\right)\right \}\nonumber\\
&=&  \int d[\lambda(\theta)] \exp\left\{- \int d\theta \lambda(\theta)\rho(\theta)\right \}
\left[\int \frac{d\theta}{2\pi}\exp \lambda(\theta)\right]^N.
\eea
We now  introduce a chemical potential $\mu$. We  multiply the last line by $(2\pi)^N \exp(N\mu)/N!$ and  sum over $N$. This gives
\be
P[\rho]=   \int d[\lambda(\theta)] \exp\left\{ \int d\theta [-\lambda(\theta)\rho(\theta)+\exp(\lambda(\theta)+\mu) ]\right\}
\ee
The value of $\mu$ will be chosen  so as to enforce $\int \rho\,d\theta = N$. In the thermodynamic limit there should be no difference between the canonical and grand canonical ensembles.

Next, the $\lambda(\theta)$ functional integral is  approximated by stationary phase. Calling the exponent $S[\lambda,\rho]$, we have
\be
\delta S= \int_0^{2\pi}  \delta \lambda(\theta) \left(-\rho(\theta)+ \exp(\lambda(\theta)+\mu) \right)\,d\theta,
\ee
Thus
\be
\lambda_{\rm stationary}(\theta)= \ln \rho(\theta) -\mu,
\ee
 and 
\be
P[\rho] \sim \exp\left\{S[\lambda_{\rm stationary},\rho]\right\}= \exp\left\{ \int_0^{2\pi} (- \rho\ln \rho  +\rho +\mu\rho)d\theta \right\}.
\ee

Corrections to the leading-order stationary-phase  result are also in powers of  $1/N$, but they have a different character from the $1/N$ corrections inherent in  $\rho\ln\rho$. The $\lambda(\theta)$ functional integral is ultra-local, and so the coefficients will involve $\delta(0)$'s \cite{jevicki1}. These divergent terms must  compensate for divergences arising in the resulting continuum $\rho(\theta)$ field theory. The underlying Schr{\"o}dinger problem, after all,  has no divergences.  

 The $-\rho\ln \rho$  in the exponent of the measure makes physical sense. It is  the configurational entropy of the non-uniform gas. The number of ways of  distributing  the $N$ distinguishable particles (they are labelled by the ``$i$'' on $\theta_i$) into $k$ bins of length $2\pi/k$, with $n_1$ in bin 1, $n_2$ in bin 2, {\it etc.\/}, is 
 \bea
 \frac{N!}{n_1! n_2!\cdots n_k!} &\approx& \exp\left\{ N\ln N- N - \sum_{\alpha=1}^k (n_\alpha \ln n_\alpha -n_\alpha)\right\} ,\nonumber\\
 &\approx& \exp\left \{ \int_0^{2\pi}(  - \rho\ln \rho +\rho+\hbox{const.})d\theta \right\} . 
 \eea
 The steepest descent approximation to the  integral over $\lambda(\theta)$ is now  seen to be the  steepest descent approximation that gives Stirling's approximation:
\bea
 \frac 1 {n !}= \frac{1}{\Gamma(n+1)} &=&\frac 1{2\pi i} \int_C t^{-(n+1)} e^{t}\,dt,\nonumber\\
 &=& \frac 1{2\pi i} \int_{C'} \exp\left\{-n \lambda +e^{\lambda}\right\}d\lambda,\nonumber\\
 &\approx& \exp\left\{ -n \ln n +n\right\}.
 \eea
In the second line  we have set $t=\exp \lambda$ and in the last line  approximated  the  integral by the  maximum value of its integrand, which occurs at $\lambda =\ln n$.

In conclusion, we have that  the Sutherland product integral
\be
\int_0^{2\pi}\ldots \int_0^{2\pi}\prod_{i=1}^N \frac{d\theta_i}{2\pi}  |\Delta|^{2\lambda}\ldots
\ee
becomes, in the collective field formalism,  proportional to a functional integral over $\rho(\theta)$ with weight \cite{andric,dyson2,polychronakos_density} 
\be
J[\rho]=\exp  \left\{( \lambda-1) \int_0^{2\pi}  \rho(\theta) \ln \rho(\theta)\, d\theta + \lambda \int_0^{2\pi} \int_0^{2\pi} \rho(\theta)\rho(\theta') \ln\left|2 \sin\left(\frac{\theta-\theta'}{2}\right) \right|\,d\theta d\theta'\right\}.
\ee
The first term in the exponent is absent in the collective-field form of the Jack inner product.

\section{Incorporating  the left-going modes}
\label{SEC:ambichiral}

In the purely right-going case the wavefunction depended only on the $p_n$ for $n$ positive, and 
$p_{-n}$ was interpreted as  the Bargmann-Fock adjoint  of the operation of  multiplication by $p_n$.  To decribe both left- and right-going  excitations simultaneously we have to allow wavefunctions   contining both  $p_n$ and $p_{-n}$. These complex variables should be  conjugates of each other, and  so the  independent variables are their  real and imaginary parts $r_n$, $s_n$ with  $n>0$.  Thus   
$p_n=r_n+is_n$  and $p_{-n}= r_n-is_n$, and 
 \be
\frac{\partial}{\partial p_n}= \frac 12 \left(\frac{\partial }{\partial r_n}-i \frac{\partial}{\partial s_n}\right),\quad 
\frac{\partial}{\partial p_{-n}}= \frac 12 \left(\frac{\partial }{\partial r_n}+i \frac{\partial}{\partial s_n}\right), \quad n>0.
\ee
Let us begin by taking  the  inner product to have the Jack-product weight 
\be
J=  \exp\left\{-\lambda \sum_{n=1}^{\infty} \frac 1n p_n p_n^*\right\}= \exp\left\{-\lambda \sum_{n=1}^{\infty} \frac 1n (r_n^2+s_n^2)\right\}.
\ee
Then, with respect to this new,  non-chiral Jack  product---let's call it Jack$'$---we  have 
\be
\left(\frac{\partial}{\partial  r_n}\right)^\dagger = -\frac 1 J\frac{\partial}{\partial  r_n} J=
-\frac {\partial}{\partial r_n} + \frac {2\lambda}{n} r_n,
\ee 
and similarly for $s_n$.
 Proceeding in this manner we find that
 \be
 \left(\frac{\partial}{\partial p_n}\right)^\dagger = -\frac{\partial}{\partial p_{-n}} +\frac\lambda {n} \sgn(n) p_{n},
 \ee
 where $n$ can have either sign. Also $p_n^\dagger =p_{-n}$. We note that $(\ldots)^{\dagger\dagger} = (\ldots)$.
 
Now  define 
 \be
 {\rm v}_n =2\pi \left(- n\frac{\partial}{\partial p_{-n}}+ \frac \lambda 2 \sgn(n) p_n\right),
 \ee
 so that  ${\rm v}_n^\dagger = {\rm v}_{-n}$ as the Jack$'$-product adjoint. With
 \be
{\rm v}(\theta)= \frac 1{2\pi}\sum_{n=-\infty}^{\infty} {\rm v}_n e^{-in\theta},\quad \rho(\theta)=  \frac 1{2\pi}\sum_{n=-\infty}^{\infty} p_n e^{-in\theta},
 \ee
 both ${\rm v}(\theta)$ and 
 $\rho(\theta)$ 
  are Jack$'$ hermitian, and 
 \be
[\rho(\theta), {\rm v}(\theta')]= - i \hbar  \partial_\theta \delta(\theta-\theta').
\ee

We now define chiral currents  $j_{\rm R,L}= \textstyle{1\over 2} (\rho\pm {\rm v}/\pi \lambda)$ with 
 \bea
 j_{{\rm R}.n}& =& -\frac{n}{\lambda} \frac{\partial }{\partial p_{-n}} +\Theta(n) p_n,\nonumber\\
j_{{\rm L}.n}& =& +\frac{n}{\lambda} \frac{\partial }{\partial p_{-n}} +\Theta(-n) p_n.\nonumber
\label{EQ:ambichiral_curent}
\eea
These obey 
\bea
{}[j_{{\rm R},n},  j_{{\rm R},m}]&=&\phantom - m\nu \delta_{m+n,0},\nonumber\\
{}[j_{{\rm L},n},  j_{{\rm L},m}]&=&-m\nu \delta_{m+n,0},\nonumber\\
{}[j_{{\rm L},n},  j_{{\rm R},m}]&=&\phantom -0,\nonumber
\eea and so the right and left current algebras are cleanly separated. We  should take   $\Theta(0)={\textstyle {1\over 2}}$ so as to agree with our previous allocation of the half of $p_0=N$ to each of  the  chiral currents. 
Unlike the chiral case, the left- and right-going  currents have $p_n$ derivatives containing {\it both \/} signs of $n$.

To write the Hamiltonian in terms of the extended set of $p_n$ we need   
\be
(z_i+z_j) \frac{z_i^{-n}-z_j^{-n}}{z_i-z_j}= - (z_i^{-n} +2 z_i^{-n+1}z_j^{-1} +\cdots +2z_i^{-1}z_j^{-n+1}  +z^{-n}_j)
\ee
in addition to our previous 
\be
(z_i+z_j)\frac{z_i^n-z_j^n}{z_i-z_j} = z_i^{n} +2z_i^{n-1}z_j+\cdots +2z_iz_j^{n-1}+z_j^{n}.
\ee
The hamiltonian becomes  \cite{gutman}
\bea
2H'&=& \sum_{n,m=-\infty}^\infty nmp_{n+m} \frac{\partial}{\partial p_n} \frac{\partial}{\partial p_m} +\lambda\sum_{n=1}^{\infty}\sum_{m=1}^{\infty}  (n+m) \left( p_{n}p_m  \frac{\partial}{\partial p_{n+m}} +
p_{-n}p_{-m}\frac{\partial}{\partial p_{-n-m}}\right)+\nonumber\\ &&+
\sum_{n=-\infty}^\infty\left((1-\lambda) n^2+\lambda |n|N\right) p_n\frac{\partial}{\partial p_{n}}.
\eea

 Now the (normal-ordered) expression 
 $$
4\pi^2 \int_0^{2\pi} \left( \frac{\lambda^2}{3}j_{\rm R}^3 +\frac{\lambda^2}{3}j_{\rm L}^3\right)d\,\theta
 $$
 is equal to 
 \bea
  \sum_{n,m=-\infty}^\infty nmp_{n+m} \frac{\partial}{\partial p_n} \frac{\partial}{\partial p_m} &+&\lambda\sum_{n=1}^{\infty}\sum_{m=1}^{\infty}  (n+m) \left( p_{n}p_m  \frac{\partial}{\partial p_{n+m}} +
p_{-n}p_{-m}\frac{\partial}{\partial p_{-n-m}}\right)\nonumber\\ &&\quad +\sum_{n=-\infty}^\infty \lambda |n|N p_n\frac{\partial}{\partial p_{n}},\nonumber
 \eea
 and  the total momentum is 
 \bea 
\hat P_{\rm tot}\equiv \sum_i D_i= \sum_{n=-\infty}^{\infty} n p_n\frac{\partial }{\partial p_n}&=&
\frac{\lambda}{2}  \sum_{n=-\infty }^\infty  \left (j_{{\rm R},n}  j_{{\rm R},-n} -  j_{{\rm L},-n}j_{{\rm L},n}\right)\nonumber\\
&=& \lambda \pi \int_0^{2\pi} (j_{\rm R}^2-j_{\rm L}^2)\,d\theta=\int_0^{2\pi} \rho {\rm v}\,d\theta.
\nonumber
\eea
In the momentum, the unwanted terms with two $\partial/\partial p_n$'s  cancelled between the left- and right-going  current contributions..

Life seems more complicated if we wish to assert that the remaining term in $2H'$
 \bea
  \sum_{n=-\infty}^{\infty} n^2 p_n\frac{\partial }{\partial p_n}&\stackrel{?}{=}&
\lambda  \sum_{n=1}^\infty n \left(j_{{\rm R},n}  j_{{\rm R},-n} +  j_{{\rm L},- n}j_{{\rm L},n}\right).\nonumber\\
 \nonumber
 \eea
 Here, although the $p_n\partial/\partial p_n$ terms are generated correctly,  the undesired two-derivative terms appearing in the right-hand-side do \underline{not}  cancel. Even worse, we find that while $\hat P_{\rm tot}$ is Hermitian thanks to  cancellations between terms with $\pm n$, the expression  $\sum  n^2 p_n{\partial }/{\partial p_n}$ is {\it not\/} Hermitian with respect to the   Jack$'$ inner product. This means that,  while in  the chiral case $H'$ was  Hermitian with respect to both the Sutherland and Jack products, in the non-chiral case  it  
is  Hermitian only with respect to the Sutherland  product.

A further  indication that the Sutherland product is essential   comes from realizing that ${\rm v}(\theta)$ is {\it not\/}  the physical velocity field. 
From number conservation 
$$
\dot \rho +\partial_\theta   \rho v=0
$$
and the Heisenberg equation of motion $\dot \rho =i[H',\rho]$,
we should have $[H',\rho]=i\partial_\theta  (\rho v)$, or, in Fourier space,
$$
n(\rho v)_n =[H,\rho_n], \quad \hbox{where}\quad  n(\rho v)_n \equiv  \frac 1{2\pi} \sum_{m=-\infty}^\infty  n p_{n-m}v_m.
$$
Now
\bea
\sum_{n,m=-\infty}^{\infty} p_n p_{m}(n+m)\sgn(m)\frac{\partial}{\partial p_{n+m}}&&\nonumber\\
=\sum_{n,m=1}^{\infty} (n+m)\left(p_np_m \frac{\partial}{\partial p_{n+m}}\right.&+&\left.p_{-n}p_{-m} \frac{\partial}{\partial p_{-n-m}}\right) + \sum_{m=-\infty}^{\infty} N|m| p_m\frac{\partial}{\partial p_m},
\label{EQ:miracle}
\eea
since
the terms with  $n$ and $m$  of opposite sign cancel, and the last term comes about because $m$ cannot be zero ($\sgn(0)=0$) but $n$ {\it can\/}  be zero, and $p_0=N$.
 Using (\ref{EQ:miracle}) we find that  
\bea
[H',\rho_n]&=& \sum_{m=-\infty}^\infty n\left(m p_{n+m} \frac{\partial}{\partial p_m}+\frac{\lambda}{2}  p_{n-m}\sgn(m) p_m\right)+ \textstyle{1\over 2} (1-\lambda)n^2 p_n\nonumber\\
&=&\frac{1}{2\pi} \sum_{m=-\infty}^\infty n p_{n-m}\left[2\pi \left(-m \frac{\partial}{\partial p_{-m}} +\frac{\lambda}{2}  \sgn(m) p_m\right)\right]+ \textstyle{1\over 2} (1-\lambda)n^2 p_n \nonumber\\
&=& \frac{1}{2\pi} \sum_{m=-\infty}^\infty n p_{n-m}{\rm v}_m  +\textstyle{1\over 2} (1-\lambda)n^2 p_n \nonumber
\eea
The first term contains our ${\rm v}_m$'s, and the last is the Fourier transform of 
$$
 i\partial_\theta  \left\{\rho\frac i2 (1-\lambda)  \partial_\theta  \ln \rho\right\}.
$$ 
We conclude that 
$$
v_{\rm physical}={\rm v}+{i\over 2}   (1-\lambda)\partial_\theta  \ln \rho.
$$
We note, however,  the comforting fact that
$$
\hat P_{\rm tot}\equiv \sum_i D_i= \int_0^{2\pi} \rho {\rm v}\,d\theta=\int_0^{2\pi} \rho v_{\rm physical}\,d\theta,
$$
because the addition to the momentum density  is a total derivative.

The distinction between ${\rm v}$ and $v_{\rm physical}$ is accounted for by the different weights in the Jack and Sutherland products.  To see this, we work in position space.
As usual we have  
$$
\rho(\theta) = \frac 1{2\pi} \sum_{n=-\infty}^{\infty} p_n e^{-in\theta}.
$$
We define
\be
\Pi(\theta)\equiv \frac {\delta}{\delta \rho(\theta)}= \sum_{n=-\infty}^{\infty}  e^{-in\theta}\frac{\partial}{\partial p_{-n}}
\ee
so  that $\delta \rho(\theta)/\delta \rho(\theta')=\delta(\theta-\theta')$. 
The operator 
 \be
-i\partial_\theta \Pi= \sum_{n=-\infty}^{\infty} \left(-n \frac {\partial}{\partial p_{-n}}\right)e^{-in\theta}
\ee 
is hermitian with respect to a product defined by an integration over $\rho$ with weight unity. If we let
\be
K[\rho]= \exp  \left\{\lambda \int_0^{2\pi} \int_0^{2\pi} \rho(\theta)\rho(\theta') \ln\left|2 \sin\left(\frac{\theta-\theta'}{2}\right) \right|\,d\theta d\theta'\right\}
\ee
be the weight appearing in the Jack product, then our 
\be
{\rm v}(\theta)  = K^{-1/2}(-i\partial_\theta  \Pi)K^{1/2} =-i\partial_\theta  \Pi - i\frac {\lambda}{2} \rho_{\rm H}
\ee
is Jack$'$ Hermitian, 
 and 
\be
v_{\rm physical} = J^{-1/2}(-i\partial_\theta  \Pi)J^{1/2} = -i\partial_\theta  \Pi - i\frac{\lambda}{2} \rho_{\rm H}- i\frac {(\lambda-1)}{2} \partial_\theta \ln \rho
\ee
is Hermitian with respect to the Sutherland product, which contains the weight $J[\rho]$.

Now   let us return to problem of expressing the remaining sum  in the Hamiltonian in terms of physical variables. We note that 
  \bea
  \sum_{n=-\infty}^{\infty} {\rm sgn}(n) n^2 p_n\frac{\partial }{\partial p_n}&{=}&
\lambda  \sum_{n=1}^\infty n \left(j_{{\rm R},n}  j_{{\rm R},-n} -  j_{{\rm L},- n}j_{{\rm L},n}\right),\nonumber\\
 \nonumber
 \eea
is Jack$'$ hermitian,  but this is not quite the expression that appears in the Hamiltonian.  We need to remove the ${\rm sgn}(n)$ by changing  the sign of the negative $n$ terms in this sum.
The following  manouvre, a paraphrase of the unusual Hilbert transform  in  \cite{abanov}, achieves this. 
We start with 
\bea
4\pi^2\int_0^{2\pi}\left\{ \frac {\lambda^2}  6 j_{\rm R}^3 +\frac {\lambda^2}6 j_{\rm L}^3\right\} d\theta  &=& \int_0^{2\pi}\left\{\frac{\lambda^2 \pi^2}6 \rho^3+\frac 12 \rho {\rm v}^2\right\}d\theta \nonumber\\
&=& \int_0^{2\pi}\left\{ \frac{\lambda^2 \pi^2}6 \rho^3+\frac 12 \rho { v}_{\rm phys} ^2 + \frac{i}{2}(\lambda-1) v_{\rm phys}\partial_\theta  \rho -\frac 18 \frac{(\partial_\theta  \rho)^2}{\rho}\right\} d\theta \nonumber
\eea
and observe  that adding   
\bea
&&\int_0^{2\pi}\left\{ \frac i 4 (\lambda-1)\left[ j_{\rm R}\partial_\theta ({\rm v}+2\pi  i\lambda \rho_{\rm H} -\pi \lambda \rho)
+j_{\rm L}\partial_\theta ({\rm v}+2\pi i\lambda \rho_{\rm H} +\pi \lambda \rho)\right]\right\}d\theta \nonumber\\
 &=&\int_0^{2\pi}\left\{ - \frac i 2(\lambda-1) v_{\rm phys} \partial_\theta  \rho + \frac 14(\lambda-1)^2 \frac{(\partial_\theta \rho)^2}{\rho} -\frac \pi 2 \lambda(\lambda-1) \rho\partial_\theta  \rho_{\rm H}\right\} d\theta\nonumber
 \eea
 gives  the  known position-space   Hamiltonian \cite{polychronakos_soliton} 
 \be
H'=\int_0^{2\pi}\left\{\frac 12 \rho v_{\rm physical}^2  +\frac{\lambda^2 \pi^2}6 \rho^3+\frac 18 (\lambda-1)^2 \frac{(\partial_\theta \rho)^2}{\rho}
 -\frac \pi 2\lambda(\lambda-1)\rho\partial_\theta   \rho_{\rm H}\right\}\,d\theta.
 \ee
 The  shift ${\rm v} \to {\rm v}+2\pi i \lambda \rho_{\rm H} \mp \pi \lambda \rho$  has changed  the signs before  the $\Theta(\pm n)$'s in the definitions of $j_{\rm L,R}$, and thus effected the desired change of sign of the negative $n$ terms in the $n^2 p_n \partial/p_n$ sum.

Setting $j_{\rm L}(\theta)=0$ in this Hamiltonian reduces it to an expression
\be
H_{\rm constrained} =4\pi^2 \int_0^{2\pi}\left\{ \frac{\lambda^2}{6}j_{\rm R}^3 - \frac{\lambda(\lambda-1)}{8\pi}  j_{\rm R}\partial( j_{\rm R})_{\rm H}\right\} d\theta. \nonumber
  \ee
 that looks very like the  chiral Hamiltonian appearing in (\ref{EQ:chiralform}). 
Further, by examining  the $n>0$ Fourier components in  (\ref{EQ:ambichiral_curent}), we see  that imposing $j_{\rm L}(\theta)=0$ as a constraint on the wavefunctions is equivalent to demanding   that  
 $$
 \frac {\partial}{\partial p_n}\to 0, \quad n<0,
 $$
 and so  requires the wavefunction not depend on $p_n$ with negative $n$. Equating the $n<0$ Fourier components to zero requires  that, as operators, we have 
 $$
 p_{-n} \to \frac n\lambda  \frac {\partial}{\partial p_n}, \quad n>0.
 $$
 Consequently,    $p_{-n} $ ceases being independent and returns to being   the Bargmann-Fock Jack-product adjoint of multiplication by $p_n$. We have precisely recovered the chiral theory from section \ref{SEC:collective}.
 The $j_{\rm L}=0$   constraint, natural as it seems,  is {\it not\/} however consistent with the full equations of motion: an initially-zero $j_{\rm L}$ does   not
 remain zero.
 
 The true, consistent, right-going chiral constraint was found by Bettelheim, Abanov and Wiegmann \cite{bettelheim} to be
\be
 v_{\rm physical}= \pi \lambda \rho -\frac 12 (\lambda-1) \partial_\theta (\ln \rho)_{\rm H}.
 \ee
 With this condition the separate continuity equation
 \be
 \dot \rho+\partial_\theta \,\rho v_{\rm physical}=0,
 \ee
 and 
 the Euler equation 
 \be
 \dot v + v_{\rm physical} \partial_\theta v_{\rm physical}= -\partial_\theta w(\rho),
 \ee
 where
 \be
 w(\rho)= \frac{\lambda^2\pi^2 \rho^2}{2}-\frac {(\lambda-1)^2}{8} \left(2\partial^2_{\theta\theta}\ln \rho+ (\partial_\theta \ln\rho)^2\right) -\pi \lambda(\lambda-1)\partial_\theta \rho_{\rm H},
 \ee
   become identical ---but only after some considerable algebra and use of Tricomi's version 
    \be
\left(\phi_1 (\phi_2)_{\rm H}\right)_{\rm H}+ \left((\phi_1)_{\rm H} \phi_2\right)_{\rm H} = (\phi_1)_{\rm H}(\phi_2)_{\rm H}-\phi_1\phi_2.
\ee
of the  Poincar{\'e}-Bertrand
    %Hardy-Tricomi 
    identity \cite{tricomi,titchmarsh}.
The resulting single equation  for the right-going wave is  \cite{bettelheim} 
\be
\dot \rho +\partial_\theta\left\{\pi \lambda \rho^2 -\frac 12 (\lambda-1) \rho \,\partial_\theta (\ln \rho)_{\rm H}\right\}=0.
 \ee 
 This equation can be made to  coincide with our earlier Benjamin-Ono equation  by   linearizing
$ \rho \,\partial_\theta (\ln \rho)_{\rm H}\approx \partial_\theta  \rho_{\rm H}$.

  In terms of the current $j_{\rm  L}$, the rather mysterious chiral condition becomes  
 \be
 \ j_{\rm L}=i\frac {(\lambda-1)}{2\pi \lambda }\partial_\theta (\ln\rho)_-
 \ee
 Recall that the subscript ``$-$'' means a projection onto  the $n<0$ Fourier modes.
Therefore, from the $n>0$ Fourier components, we again read off that 
 \be
 \frac{\partial}{\partial p_{n}}\to 0, \quad n<0,
 \ee
 and  the wavefunction remains only a function of the $p_n$ for $n>0$. 
 The $n<0$ components, however, now  give
 \be
p_{-n} \to \frac{n}{\lambda} \frac {\partial}{\partial p_n}+  i\frac {(\lambda-1)}{2\pi \lambda }\left [\partial_\theta (\ln\rho)\right]_{-n}, \quad n>0.
 \ee
 This equation  asserts   that  that $p_{-n}=p_n^\dagger$ with the adjoint taken with respect to the  {\it Sutherland\/} product.  The true chiral condition is therefore a very natural, and indeed inevitable, consequence of the necessity of using only the Sutherland inner product  when dealing with both the full ambichiral collective field.
 
 \section{Conclusions}
 \label{SEC:conclusions}
 
 We have traced the difficulty in separating the left- and right-going degrees of freedom in the continuum hydrodynamics of the Sutherland model to the existence of two distinct inner products with respect to which the polynomial  eigenfunctions are orthogonal.   Each chiral half of the model is most naturally expressed in terms of operators that are hermitian with respect to the first  of these products, but the full model is only hermitian with respect to the second. 
 
 We have still not managed to decouple the oppositely moving edge modes into non interacting waves, and it is an interesting question whether this is possible.

 \section{Acknowledgements}

We thank A.~Abanov and P.~Wiegmann for explaining  how they think of the chiral condition, and for useful comments on the manuscript.  MS would also like to thank Inaki Anduaga and Lei Xing for discussions and help.  Work in Urbana was supported by the National Science Foundation under grant DMR-06-03528, and work in Gainsville under grant DMR-03-08377.

\end{document}